\documentclass[article,aps,nofootinbib,twocolumn,superscriptaddress]{revtex4-1}
\input epsf
\usepackage{graphics}
\usepackage{amsmath}
\usepackage{amssymb}
\usepackage{bm}
\usepackage{color}
\usepackage{dcolumn}
\usepackage{hyphenat}

\allowdisplaybreaks

\def\be{\begin{equation}}
\def\ee{\end{equation}}
\def\ba{\begin{eqnarray}}
\def\ea{\end{eqnarray}}

\usepackage{graphicx}

\begin{document}

\title{What can Cosmology tell us about Gravity? \\ Constraining Horndeski with $\Sigma$ and $\mu$}

\author{Levon Pogosian} \affiliation{Department of Physics, Simon Fraser University, Burnaby, BC, V5A 1S6, Canada}
\author{Alessandra Silvestri} \affiliation{Institute Lorentz, Leiden University, PO Box 9506, Leiden 2300 RA, The Netherlands}

\begin{abstract}
Phenomenological functions $\Sigma$ and $\mu$ (also known as $G_{\rm light}/G$ and $G_{\rm matter}/G$) are commonly used to parameterize possible modifications of the Poisson equation relating the matter density contrast to the lensing and the Newtonian potentials, respectively. They will be well constrained by future surveys of large scale structure. But what would the implications of measuring particular values of these functions be for modified gravity theories? We ask this question in the context of general Horndeski class of single field scalar-tensor theories with second order equations of motion. We find several consistency conditions that make it possible to rule out broad classes of theories based on measurements of $\Sigma$ and $\mu$ that are independent of their parametric forms. For instance, a measurement of $\Sigma \ne 1$ would rule out all models with a canonical form of kinetic energy, while finding $\Sigma-1$ and $\mu -1$ to be of opposite sign would strongly disfavour the entire class of Horndeski models. We separately examine the large and the small scale limits, the possibility of scale-dependence, and the consistency with bounds on the speed of gravitational waves. We identify sub-classes of Horndeski theories that can be ruled out based on the measured difference between $\Sigma$ and $\mu$.
\end{abstract}

\maketitle

\section{Introduction}

General Relativity (GR) \cite{Einstein:1915:FGG} provides a theoretical framework for calculating predictions of cosmological models and testing them against observations made on the sky. As the variety and the quality of observations improve, it is becoming possible to not only test particular models within the framework of GR but, in addition, to test the consistency of GR itself~\cite{Ishak:2005zs,Linder:2005in,Bertschinger:2006aw,Zhang:2007nk,Amendola:2007rr}. Aside from the new opportunities for testing gravity on cosmological scales, the observed cosmic acceleration \cite{Perlmutter:1998np,Riess:1998cb} and the unexplained nature of Dark Matter led to an increased interest in alternative gravity theories (for reviews, see \cite{Silvestri:2009hh,Clifton:2011jh,Joyce:2014kja}). Additional motivation comes from the long standing failure to explain the technically unnatural fine-tuning needed to reconcile the very large vacuum energy predicted by particle physics with the small value of the observed cosmological constant \cite{Weinberg:1988cp,Burgess:2013ara}.

Significant amount of work over the past decade went into understanding the aspects of GR that can be tested observationally and developing frameworks and practical tools for implementing these tests \cite{Hu:2007pj,Bertschinger:2008zb,Zhao:2008bn,Pogosian:2010tj,Hojjati:2011ix,Baker:2012zs,Battye:2012eu,Gubitosi:2012hu,Bloomfield:2012ff,Hu:2013twa,Gleyzes:2013ooa,Bellini:2014fua,Gleyzes:2014rba,Baker:2014zva,Zumalacarregui:2016pph}. The validity range of such frameworks is generally restricted to linear cosmological scales. Much like the Parameterized Post-Newtonian formalism \cite{Nordtvedt:1968qr,Nordtvedt:1968qs,Will:1971zzb,Will:1972zz}, they involve phenomenological parameters or functions that can be constrained and compared to predictions of specific theories. 

One of the testable aspects of GR is the relationship between the curvature perturbation $\Phi$ and the Newtonian potential $\Psi$. In GR, when the matter anisotropic stress can be neglected,  the Weyl potential, $\Phi_+ \equiv (\Phi+ \Psi)/2$, affecting relativistic particles, is the same as the gravitational potential felt by non-relativistic particles. In contrast, alternative gravity theories typically contain additional degrees of freedom that can mediate new interactions. There, the equivalence between $\Phi_+$, $\Phi$ and $\Psi$ is generically broken. By combining  weak lensing and galaxy clustering data from the upcoming large scale structure surveys, such as Euclid and LSST, one can search for differences between the different potentials and constrain alternative gravity theories. 

While $\Phi \ne \Psi$ is a generic signature of a non-minimal gravitational coupling, the quantities that will be more directly probed by observations of galaxy redshifts and weak lensing are the effective gravitational constants $G_{\rm matter}$ and $G_{\rm light}$ that appear, respectively, in the Poisson equations for $\Psi$ and $\Phi_+$. Parameters $\Sigma = G_{\rm light}/G$ and $\mu=G_{\rm matter}/G$, which generally are functions of scale and redshift, have been widely used in papers on cosmological tests of GR \cite{Zhao:2010dz,Song:2010fg,Hojjati:2011xd,Simpson:2012ra,Ade:2015rim,Blake:2015vea}. Another widely used parameter, $E_G$, introduced in \cite{Zhang:2007nk}, is designed to directly probe the relation between $\Phi_+$ and $\Psi$.  As shown in \cite{Leonard:2015cba}, practical implementations of the $E_G$ test \cite{Reyes:2010tr,Blake:2015vea} are, in effect, primarily sensitive to $\Sigma$. 

Despite the high sensitivity of observables to $\Sigma$ and $\mu$, and their widespread use as phenomenological parameters, the physical implications of measuring $\Sigma \ne 1$ or $\mu \ne 1$ have not been fully explored. As we argue in this paper, the measurement of $\Sigma$, and its difference from $\mu$, are of key importance for discriminating among modified gravity theories. For instance, scalar-tensor theories with a scalar that has a canonical kinetic term, {\it i.e.} the generalized Brans-Dicke (GBD) models, predict a scale-independent $\Sigma$ which is strongly constrained to be close to unity in models with universal coupling to different matter species. Thus, a measurement of $\Sigma \ne 1$ would rule out all universally coupled GBD theories, such as the $f(R)$ \cite{Capozziello:2003tk,Carroll:2003wy,Appleby:2007vb,Hu:2007nk,Starobinsky:2007hu}, chameleon \cite{Khoury:2003aq}, symmetron \cite{Hinterbichler:2010es} and dilaton \cite{Damour:1994zq,Brax:2011ja} models. 

\begin{figure}
\centering
\includegraphics[width=0.5\textwidth]{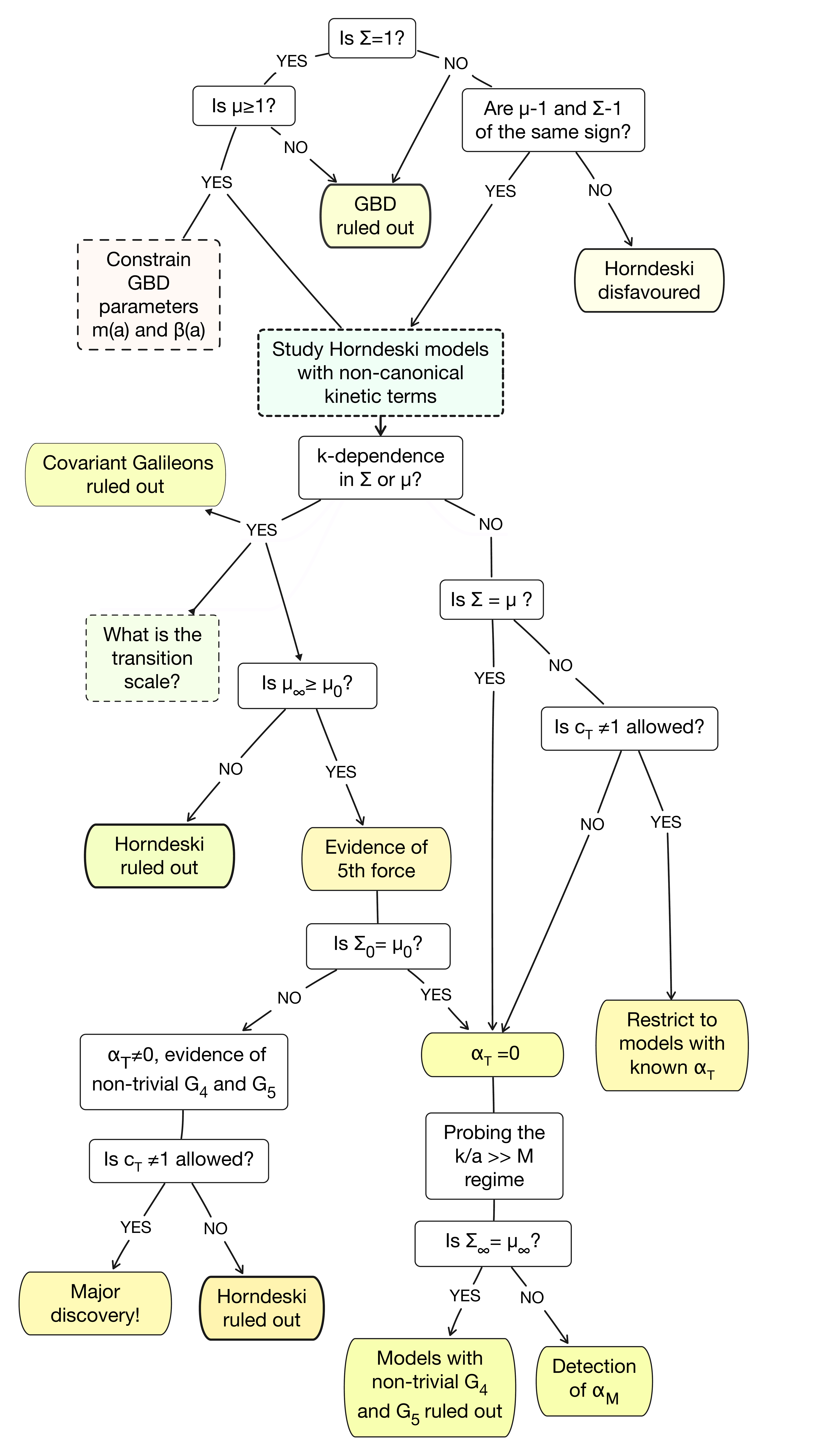}
\caption{A flow chart diagram summarizing the main conclusions of the paper. It provides a systematic way of interpreting measured values of phenomenological functions $\Sigma$ and $\mu$ for the purpose of constraining, and even ruling out, scalar-tensor theories of Horndeski type. Note that, while not explicitly indicated on the diagram, measuring any of the functions to be different from 1 would rule out GR. Also, measuring $\Sigma=\mu=1$ would imply consistency of observations with GR as well as with Horndeski theories, since the latter includes GR.}
\label{fig:chart}
\end{figure}

Within a broader class of models, such as the Horndeski class \cite{Horndeski:1974wa,Deffayet:2011gz,Kobayashi:2011nu} of general scalar-tensor theories with up to second order equations of motion, a number of useful results were obtained in \cite{Bellini:2014fua,Tsujikawa:2015mga,Perenon:2015sla,Lombriser:2015sxa,Gleyzes:2015rua,Salvatelli:2016mgy,Lombriser:2016yzn} concerning general features of the growth of structure in the quasi-static limit. Our aims are similar to those in \cite{Tsujikawa:2015mga,Perenon:2015sla,Gleyzes:2015rua}, and our conclusions agree where they overlap, but the questions we address are more specifically focused on the implications of measuring particular values of  $\mu$ and $\Sigma$ for sub-classes of scalar-tensor models. We avoid making strong theoretical assumptions about the anomalous speed of gravity waves, and do not require the models to allow for self-acceleration, since Dark Energy is only one possible motivation for studying modifications of gravity. We also avoid making assumptions about particular functional forms of the free functions or priors on their parameters, focusing instead on trends that are parametrization independent. The main conclusions of this paper are presented in the form of a flow chart diagram in Fig.~\ref{fig:chart}.

In what follows, we review the definition of the phenomenological functions $\Sigma$, $\mu$ and $\gamma$ in Sec.~\ref{sec:definitions} and the basics of the effective theory approach to linear perturbations in scalar-tensor models in Sec.~\ref{sec:st}. In Sec.~\ref{sec:horndeski}, we examine the expressions for the phenomenological functions in the quasi-static limit in Horndeski models and observe several consistency relations that can be tested with observations. We consider some examples in Sec.~\ref{sec:examples} and conclude with a summary in Sec.~\ref{sec:summary}.  One of our conclusions is that measuring $\Sigma - 1$ and $\mu - 1$ to be of opposite sign would effectively rule out all Horndeski models.

\section{The phenomenological functions $\Sigma$, $\mu$ and $\gamma$.}
\label{sec:definitions}

Consider the perturbed Friedmann-Lemaitre-Robertson-Walker (FLRW) metric in the conformal Newtonian gauge
\be
ds^2 = -(1+2\Psi)dt^2 + a^2(1-2\Phi) d{\bf x}^2 \ ,
\ee
where $a$ is the scale factor. Einstein's equations of General Relativity relate potentials $\Psi$ and $\Phi$ to the components of the perturbed stress-energy tensor. Specifically, working in Fourier space, one can combine the $00$ and the $0i$ components of the Einstein equations to form the Poisson equation
\be
k^2 \Phi = -4\pi G a^2 \rho \Delta \ ,
\label{poisson_gr}
\ee
while the $i \ne j$ component gives
\be
k^2(\Phi -\Psi) = 12 \pi G a^2 (\rho+P) \sigma \ ,
\label{shear_gr}
\ee
where ${\bf k}=\hat{k} k$ is the Fourier vector, $G$ is the gravitational constant, $\rho$ is the background matter density, $\Delta$ is the comoving density contrast and $\sigma$ is the dimensionless shear perturbation\footnote{$\Delta \equiv \delta + 3aHv/k$, where $\delta \equiv \delta \rho/\rho$ is the density contrast in the Newtonian conformal gauge, $v$ is the irrotational component of the peculiar velocity, $H=\dot{a}/a$; $(\rho+P)\sigma \equiv -(\hat{k}^i \hat{k}_j - \delta^i_j/3))\pi^j_i$, where $\pi^j_i$ is the traceless component of the energy-momentum tensor.}. 

Eqs.~(\ref{poisson_gr}) and (\ref{shear_gr}) can be combined into an equation relating the Weyl potential, $\Phi_+ \equiv (\Phi +\Psi)/2$, to the stress-energy components:
\be
2k^2 \Phi_+ = k^2(\Phi +\Psi) = -8\pi G a^2 [ \rho \Delta + 3(\rho+P) \sigma/2] \ .
\label{weyl_gr}
\ee
Non-relativistic particles respond to gradients of the gravitational potential $\Psi$, while relativistic particles ``feel'' the gradients of the Weyl potential $\Phi_+$. In LCDM, at epochs when radiation density can be neglected, $\sigma=0$ and one has $\Phi_+ = \Phi = \Psi$. However, in alternative models, in which additional degrees of freedom can mediate gravitational interactions, the three potentials need not be equal. It will be possible to test this by combining the weak lensing shear and galaxy redshift data from surveys like Euclid and LSST. A common practical way of conducting such tests involves introducing phenomenological functions $\mu$, $\gamma$ and $\Sigma$, parameterizing departures of Eqs.~(\ref{poisson_gr}), (\ref{shear_gr}) and (\ref{weyl_gr}) from their LCDM form. Neglecting the radiation shear ($\sigma=0$), which is irrelevant at epochs probed by the surveys, they are defined as
\ba
 k^2 \Psi &=& -4\pi G \mu (a,k) a^2 \rho \Delta \ ,
\label{poisson_mg} \\
 \Phi &=& \gamma(a,k) \Psi \ ,
\label{shear_mg} \\
 k^2(\Phi +\Psi) &=& -8\pi G\, \Sigma(a,k)\, a^2 \rho \Delta \ .
\label{weyl_mg}
\ea
The three functions are related, so providing any two of them is sufficient for solving for the evolution of cosmological perturbations~\cite{Pogosian:2010tj} as, for example, implemented in the publicly available code MGCAMB \cite{Zhao:2008bn,Hojjati:2011ix}.

In general,  cosmological perturbations can be solved for exactly on \emph{all} linear scales, once two of the above functions are provided. However, deriving the functional forms of these functions in a specific gravity theory requires taking the quasi-static approximation (QSA). Under the QSA, one restricts to scales below the sound horizon of the scalar field and ignores time-derivatives of the gravitational potentials and the scalar field perturbations. We discuss this further in Sec.~\ref{sec:qsa}.

Detailed principal component analysis forecasts for surveys like LSST and Euclid \cite{Hojjati:2011xd,Asaba:2013mxj} show that $\Sigma$ is the parameter that is best constrained by the combination of weak lensing and photometric galaxy counts. Adding information from measurements of redshift space distortions, afforded with spectroscopic galaxy redshifts, adds a bias-free estimate of the Newtonian potential and helps to further break the degeneracy between $\Sigma$ and $\mu$ \cite{Song:2010fg,Simpson:2012ra,Asaba:2013mxj}. The parameter $\gamma$ is generally more weakly constrained \cite{Zhao:2009fn,Hojjati:2011xd,Asaba:2013mxj} because it is not directly probed by the observables and is effectively derived from the measurement of the other two. From the physical perspective, it is informative to examine constraints on $\mu$, $\gamma$ and $\Sigma$ simultaneously,  because specific models predict consistency relations among them.

\section{Scalar-tensor theories}
\label{sec:st}

Essentially all attempts to modify GR result in theories with additional degrees of freedom \cite{Weinberg:1965rz,Lovelock:1971yv}.
Even when these degrees of freedom are not fundamental scalar fields, they can manifest themselves as such in limits appropriate for cosmological structure formation (see \cite{Clifton:2011jh} for a review of proposed alternative gravity models).

In this Section, we review the effective theory approach to scalar-tensor models of gravity that covers linear perturbations in all single scalar field models.

\subsection{The effective approach to Dark Energy}

The effective theory (known as ``EFT'' or ``Unified'') approach to Dark Energy  \cite{Gubitosi:2012hu,Bloomfield:2012ff,Gleyzes:2013ooa,Bloomfield:2013efa} provides a unifying language for studying the dynamics of linear perturbations in the broad range of single scalar field models of dark energy and modified gravity. This includes the Horndeski class~\cite{Horndeski:1974wa},  beyond Horndeski models, such as those of~\cite{Gleyzes:2014qga}, as well as the ghost condensate model~\cite{ArkaniHamed:2003uy} and low energy versions of Lorentz violating theories like Ho\v rava-Lifshitz gravity~\cite{Horava:2009uw,Kase:2014cwa,Frusciante:2015maa}. Inspired by the EFT of Inflation~\cite{Cheung:2007st}, it is based on writing an action for the perturbed FLRW metric that includes all terms invariant under time-dependent spatial diffeomorphisms up to the quadratic order in perturbations. The action is constructed in the unitary gauge, in which the slices of constant time are identified with the hypersurfaces of uniform scalar field\footnote{Our analysis concerns cosmological perturbations around the FRW background that can be probed with large scale surveys. For complementary analysis of perturbations in scalar-tensor theories around different backgrounds, such as spherical symmetry, see for instance \cite{Kobayashi:2012kh,Kase:2014baa}.}. It is assumed that all matter fields minimally couple to the same Jordan frame metric, however, in principle, one could relax this assumption and allow for different couplings~\cite{Gleyzes:2015pma}. The resulting EFT action can be written as
\ba
\nonumber
 S &=& \int d^4x \sqrt{-g} \bigg\{ \frac{m_0^2}{2} \Omega(t) R+ \Lambda(t) - c(t) \delta g^{00} \\ \nonumber
 &+& \frac{M_2^4 (t)}{2} (\delta g^{00})^2 - \frac{\bar{M}_1^3 (t)}{2} \delta g^{00} \delta K^\mu_\mu
 - \frac{\bar{M}_2^2 (t)}{2} (\delta K^\mu_\mu)^2 \\ \nonumber
    &-& \frac{\bar{M}_3^2 (t)}{2} \delta K^i_j \delta K^j_i
    + \frac{\hat{M}^2 (t)}{2} \delta g^{00} \delta R^{(3)} \\ \nonumber
   &+& m_2^2(t)\left(g^{\mu\nu}+n^{\mu} n^{\nu}\right)\partial_{\mu}(g^{00})\partial_{\nu}(g^{00})   \bigg\} \\
 &+& S_{m} [g_{\mu \nu}, \chi_i] \ ,
\label{EFT_action}
\ea
where $m_0^{-2} = 8\pi G$, and $\delta g^{00}$, $\delta {K}{^\mu_\nu}$, $\delta K$ and $\delta R^{(3)}$ are, respectively, the perturbations of the time-time component of the metric, the extrinsic curvature and its trace, and the three dimensional spatial Ricci scalar of the constant-time hypersurfaces. Finally, $S_m$ is the action for all matter fields $\chi_i$ minimally coupled to the metric $g_{\mu \nu}$. 

The perturbation of the scalar field can be made explicit applying an infinitesimal time-diffeomorphism, $t\rightarrow t+\pi(x^\mu)$. This restores covariance and  action~(\ref{EFT_action}) is then written in terms of coordinates defined on hypersurfaces of constant background density, with $\pi$ representing the perturbed part of the scalar degree of freedom. After performing this transformation, one can choose the Newtonian gauge and study linear growth of structure in the standard way. The corresponding Poisson and  anisotropic shear equations can be written as~\cite{Bloomfield:2012ff}
\ba\label{Poisson}
&&2m_0^2 \Omega {k^2 \over a^2} \Phi = -\rho \Delta + \Delta_P\\
\label{anisotropy}
&&m_0^2 \Omega {k^2 \over a^2} (\Phi - \Psi)= {3\over 2} (\rho+P) \sigma + \Delta_S
\ea
where $\Delta_P$ and $\Delta_S$ denote the additional terms in the Poisson and shear equations. Under the QSA, they read
\ba
{a^2 \over k^2} \Delta_P= &&(m_0^2{\dot \Omega} + {\bar M}^3_1 - 2H  {\bar M}^2_3  - 4H  {\hat M}^2) \pi
\nonumber \\ 
&&- 4H  {\hat M}^2 \Phi + 8m^2_2 \Psi,
\\
{a^2 \over k^2} \Delta_S = &&(m_0^2{\dot \Omega}  - {\bar M}^2_3 H - 2{\bar M}_3 \dot {\bar M}_3 ) \pi +  2{\hat M}^2 \Psi\,.
\ea

For models with a canonical form of the scalar field kinetic energy, with the Lagrangian given by Eq.~(\ref{gbd_lagrangian}) in Sec.~\ref{sec:gbd}, all coefficients in the EFT action (\ref{EFT_action}) are zero except for $\Omega$, $\Lambda$ and $c$.  In that case,  
\be
\Delta_P=\Delta_S = m_0^2{\dot \Omega} \pi k^2/a^2, 
\ee
and one immediately finds $\Sigma=1/\Omega$ after subtracting (\ref{anisotropy}) from (\ref{Poisson}).

To derive $\Sigma$ in a general scalar-tensor theory, as well as the expressions for the other phenomenological functions $\mu$ and $\gamma$, one needs to supplement Eqs.~(\ref{Poisson}) and (\ref{anisotropy}) with the scalar field equation of motion so that $\pi$ can be eliminated from the system of equations. The general expressions for $\Sigma$, $\mu$ and $\gamma$ that follow are lengthy and not particularly illuminating, thus we opt to show them in the Appendix.

In what follows, we specialize to the Hordneski sub-class of scalar-tensor theories, which cover all models with manifestly second order equations of motion. 

\subsection{Horndeski models and their ``effective'' representation}

The most general action for a scalar-tensor theory in (3+1) dimensions with second-order field equations was originally written by Horndeski \cite{Horndeski:1974wa} and, more recently, rediscovered in \cite{Deffayet:2011gz,Kobayashi:2011nu} in the context of generalized Galileon models. The Horndeski action is given by
\begin{equation}
S = \int d^4x \sqrt{-g} \left[\sum_{i=2}^{5}{\cal L}_{i} + {\cal L}_M(g_{\mu \nu},\psi) \right] \ ,
\label{S_Horndeski}
\end{equation}
with
\begin{eqnarray}
\nonumber
{\cal L}_{2} & = & K(\phi,X),\\ \nonumber
{\cal L}_{3} & = & -G_{3}(\phi,X)\Box\phi,\\ \nonumber
{\cal L}_{4} & = & G_{4}(\phi,X)\, R+G_{4X}\,[(\Box\phi)^{2}-(\nabla_{\mu}\nabla_{\nu}\phi)\,(\nabla^{\mu}\nabla^{\nu}\phi)]\,,\\ \nonumber
{\cal L}_{5} & = & G_{5}(\phi,X)\, G_{\mu\nu}\,(\nabla^{\mu}\nabla^{\nu}\phi) \\ \nonumber
&-&\frac{1}{6}\, G_{5X}\,[(\Box\phi)^{3}-3(\Box\phi)\,(\nabla_{\mu}\nabla_{\nu}\phi)\,(\nabla^{\mu}\nabla^{\nu}\phi) \\
&+&2(\nabla^{\mu}\nabla_{\alpha}\phi)\,(\nabla^{\alpha}\nabla_{\beta}\phi)\,(\nabla^{\beta}\nabla_{\mu}\phi)] \ ,
\label{Horn_Li}
\end{eqnarray}
where $K$ and $G_{i}$ ($i=3,4,5$) are functions of the scalar field $\phi$ and its kinetic energy $X=-\partial^{\mu}\phi\partial_{\mu}\phi/2$, $R$ is the Ricci scalar, $G_{\mu\nu}$ is the Einstein tensor, and $G_{iX}$ and $G_{i\phi}$ denote the partial derivatives of $G_{i}$ with respect to $X$ and $\phi$, respectively.

For the Horndeski class of models,
\be
m_2^2=0; \ 2{\hat M}^2 = {\bar M}^2_2 =-{\bar M}^2_3,
\ee
and the relations between the  EFT functions $\Omega$, $\Lambda$, $c$, $\bar{M}^2_1$, $M_2^4$, ${\hat M}^2$ appearing in (\ref{EFT_action}) and the functions in the Horndeski Lagrangian (\ref{Horn_Li}) can be found in \cite{Bloomfield:2013efa}.

An equivalent alternative way of parameterizing the EFT action for linear perturbations around a given FLRW background in Horndeski models was introduced in \cite{Bellini:2014fua,Gleyzes:2014rba}:
\ba
\nonumber
 S^{(2)} &=& \int dtdx^3\, a^3 {M_*^2 \over 2} \bigg\{ \delta K^i_j \delta K^j_i - \delta K^2 + R \delta N \\ \nonumber
 &+& (1+ \alpha_T) \delta_2 \left( \sqrt{h}R/a^3 \right) + \alpha_K H^2 \delta N^2 \\
 &+& 4\alpha_B H \delta K \delta N  \bigg\}+ S^{(2)}_{m} [g_{\mu \nu}, \chi_i] \ ,
\label{alpha_action}
\ea
where $N$ is the lapse function and $S^{(2)}_{m}$ is the action for matter perturbations in the Jordan frame. This action is parameterized by five functions of time: the Hubble rate $H$, the generalized Planck mass $M_*$, the gravity wave speed excess $\alpha_T$, the ``kineticity'' $\alpha_K$, and the ``braiding'' $\alpha_B$ \cite{Bellini:2014fua}. It is also convenient to define a derived function, $\alpha_M$, which quantifies the running of the Planck mass. For known solutions of the Horndeski theories, they can be expressed in terms of the functions appearing in the Lagrangian (\ref{Horn_Li}), with the relations provided in Appendix~\ref{sec:eom-eft}. There are notable connections between these effective functions and the phenomenology of Horndeski theories:
\begin{itemize}
\item $\alpha_T=c_T^2-1$ is the excess speed of gravity waves, and is non-zero whenever there is a non-linear derivative coupling of the scalar field to the metric. The same non-linearity is responsible for a non-zero anisotropic stress component in the scalar field energy-momentum tensor.
\item $\alpha_K$ quantifies the ``independent'' dynamics of the scalar field, stemming from the existence of a kinetic energy term in the scalar field Lagrangian. For example, $\alpha_K\ne 0$ in minimally coupled scalar fields, such as quintessence and k-essence, while $f(R)$ models have $\alpha_K=0$. In the latter case, the scalar field is $df/dR$, and is completely determined by the dynamics of the Ricci scalar.
\item $\alpha_B$ signifies a coupling between the metric and the scalar field degrees of freedom. It is zero for minimally coupled models, such as quintessence and k-essence, and non-zero for all known modified gravity models, {\it i.e.} all models with a fifth force.
\item The running of the Planck mass, $\alpha_M$, is also generated by a non-minimal coupling, but of a more restricted type. All known models with $\alpha_M \ne 0$, also have $\alpha_B \ne 0$, but the reverse statement is not true. E.g. for $f(R)$, $\alpha_M = - \alpha_B$, while in the ``kinetic gravity braiding'' model \cite{Deffayet:2010qz}, one has $\alpha_B \ne 0$ and $\alpha_M = 0$.
\end{itemize}

It is also interesting to note the connection between the higher order derivative terms in the Horndeski action, the scalar field anisotropic stress, and the speed of gravity waves \cite{Saltas:2014dha}. A non-vanishing function ${\bar M}^2_2 =2{\hat M}^2 = -{\bar M}^2_3$ generates a non-zero shear component of the scalar field energy-momentum \cite{Bloomfield:2012ff}. Such a component does not exist for canonical scalar fields and originates from the non-linearities generated by higher order kinetic energy terms. The same non-linearity is responsible for the modification of the dispersion relation for gravitational waves \cite{Gao:2011qe,DeFelice:2011bh}, leading to a change in $c_T^2$. From the mapping provided in Appendix \ref{sec:eom-eft}, we have
\ba
&&2H M_*^2 \alpha_T = {\bar M}^2_2 =\nonumber \\ 
&& 2X[2G_{4X}-2G_{5\phi}-(\ddot{\phi} - H\dot{\phi}) G_{5X}],
\label{eq:m_bloom}
\ea
{\it i.e.} the anisotropic stress and $\alpha_T$ are non-zero if either $G_{4X}$, $G_{5\phi}$ or $G_{5X}$ is not zero. 

Both ways of parameterizing the effective action,  (\ref{EFT_action}) and (\ref{alpha_action}), have their merits, and one or the other can be preferred depending on the circumstances. As mentioned before,  (\ref{EFT_action}) was designed to cover a broader range of models, while (\ref{alpha_action}) is optimized to Horndeski but can be extended to ``beyond Horndeski'' \cite{Gleyzes:2015rua} and other models \cite{Frusciante:2016xoj}. Also, (\ref{EFT_action}) simultaneously parameterizes the evolution of the background and the perturbations and, as such, is more directly related to the full Lagrangian of particular models. For example, both the background and the perturbations in the entire class of GBD models can be described by specifying two functions of time, $\Omega(t)$ and $\Lambda(t)$, that have transparent physical meanings of, respectively, the conformal coupling and the difference between the scalar field kinetic and potential energy densities. In contrast, doing the same in the framework of (\ref{alpha_action}) requires specifying four functions: $H(t)$, $\alpha_K(t)$, $\alpha_B(t)$ and $\alpha_M(t)$, with an obscured connection to the original Lagrangian. On the other hand, working with $\alpha$'s is more efficient in an agnostic approach to testing general Horndeski models, as demonstrated in the Section below.

\section{Phenomenology of Horndeski}
\label{sec:horndeski}

In this Section, we examine the forms of the phenomenological functions $\Sigma$, $\mu$ and $\gamma$ in Horndeski models. We consider the scale-dependence associated with the mass of the scalar degree of freedom, $M$, and examine the limiting cases of $k/a \ll M$ and $k/a \gg M$, since the range of linear scales actually probed by observations is likely to fall into one of these two regimes. We point out consistency checks that can help to determine which of the two limiting regimes happened to fall into the observational window, as well as tests that can be performed if the $k$-dependence is detected. We also briefly address the conditions for validity of the QSA.

\subsection{The Compton transition scale}

Scalar-tensor theories have a scale associated with the Compton wavelength of the scalar field that sets the range of the fifth force. It is determined by the ``mass term'' term (the $C_\pi$ term) in the equation of motion for the scalar field perturbations, given in Eq.~(\ref{eom_pi}) of the Appendix.

The expressions for the phenomenological functions $\mu$, $\gamma$ and $\Sigma$ in Horndeski theories can be readily obtained from Eqs.~(\ref{eq:mu-eft}), (\ref{eq:gamma-eft}), and (\ref{eq:Sigma-eft}) of the Appendix. As we are specifically interested in the scale dependence, we can write them as
\ba
\mu &=& {m_0^2 \over M_*^2} {1+ M^2 \ a^2/k^2 \over f_3/2f_1M_*^2+M^2(1+\alpha_T)^{-1} a^2/k^2},
\label{eq:mu-horn} \\
\gamma &=& {f_5/f_1+M^2(1+\alpha_T)^{-1} \ a^2/k^2 \over 1+ M^2 \ a^2/k^2},
\label{eq:gamma-horn} \\
\Sigma &=& {m_0^2 \over 2M_*^2} {1+f_5/f_1+M^2[1+(1+\alpha_T)^{-1}] a^2/k^2 \over f_3/2f_1M_*^2+M^2(1+\alpha_T)^{-1} a^2/k^2},
\label{eq:Sigma-horn}
\ea
where we defined $M^2 \equiv C_\pi /f_1$ and used Eqs.~(\ref{eq:M*}) and (\ref{eq:alphaT}) to convert to the notation in (\ref{alpha_action}). From (\ref{eq:mu-horn}), (\ref{eq:gamma-horn}) and (\ref{eq:Sigma-horn}), one can see that $M^2$ sets the transition scale in all three phenomenological functions. The differences amount to factors of $(1+\alpha_T)$ which, as we discuss in the next subsection, are constrained to be close to unity.

For most of the specific models studied in the literature, the observational window offered by surveys of large scale structure happens to be either entirely below or entirely above the Compton wavelength. For instance, in models that exhibit self-acceleration, such as Covariant Galileons, the scalar mass is very small, comparable to $H$. Then, as far as the large scale structure observables are concerned, one always probes the small scale regime, $k/a \gg M $. Thus, \emph{a detection of $k$-dependence in either $\Sigma$ or $\mu$ would rule out self-accelerating models, such as Covariant Galileons.}

Self-acceleration is not the only motivation for studying modifications of gravity, and one could have scalar fields of larger masses that mediate new interactions without providing an alternative to Dark Energy. Popular examples include GBD models of chameleon type, in which the Compton wavelength is constrained to be $\lesssim 1$Mpc. With this in mind, we will consider both the large and the small scale limits, as well as the possibility of transition occurring inside the observational window, and try to identify testable consistency relations.

\subsection{The large scale limit}

Taking the $k/a \ll M$ limit in Eqs.~(\ref{eq:mu-horn}), (\ref{eq:gamma-horn}) and (\ref{eq:Sigma-horn}), we obtain
\ba
\mu_0 &=&{m_0^2 \over M_*^2}(1+\alpha_T), \label{mu0_Hornd} \\
\gamma_0 &=& {1 \over 1+\alpha_T} = c_T^{-2}, \label{gamma0_Hornd}\\
\Sigma_0 &=& {m_0^2 \over M_*^2} \left(1+{\alpha_T \over 2} \right). 
\label{Sigma0_Hornd}
\ea
Thus, on large scales, modifications are either due to change in the background value of the Planck mass, or due to the modified propagation speed of gravity waves. Note that both are independent of the fluctuations in the scalar field, since the scalar fifth force is suppressed on scales above the Compton wavelength. 

The \emph{current} effective value of $G_{\rm matter}$ in the $k \rightarrow 0$ limit must coincide with the value measured in Cavendish type experiments on Earth \cite{Perenon:2015sla}. Even though Earth must be in a screened environment to satisfy the stringent laboratory and solar system tests of GR, screening mechanisms of chameleon \cite{Khoury:2003aq,Hinterbichler:2010es,Damour:1994zq,Brax:2011ja} or Vainshtein \cite{Vainshtein:1972sx} type only suppress the enhancement in the effective Newton's constant caused by the attractive force mediated by scalar field fluctuations. They do not affect the super-Compton value of the gravitational coupling. This imposes a constraint on the current values of $\mu_0$ and $\Sigma_0$:
\be
\mu_0(t_0)=1, \ \Sigma_0(t_0)= {1\over 2} \left( 2+\alpha_T \over 1+ \alpha_T \right) \ ,
\ee
with $\gamma_0$ still given by (\ref{gamma0_Hornd}). Bounds on the current value of the gravity wave speed come from non-observation of the gravitational Cherenkov radiation by cosmic rays~\cite{Caves:1980jn,Moore:2001bv}. This strongly constrains the possibility of $c_T<1$, or $\alpha_T<0$. However, as argued in \cite{Jimenez:2015bwa}, in principle, the speed of the extremely high energy gravitons ($\sim 10^{10}$eV) involved in deriving this bound needs not necessarily be the same as the propagation speed of linear tensor mode metric perturbations, given the non-linearity of the coupling of the metric to the scalar field. Another bound, derived in \cite{Jimenez:2015bwa}, comes from the observed evolution of the orbital period of binary pulsars, constraining $\alpha_T$ to be within $10^{-2}$. Tight direct bounds on $c_T$ will become available when the electromagnetic counterparts of the gravity wave emitting events are detected at cosmological redshifts \cite{Nishizawa:2014zna}.

While the present value of $\alpha_T$ is strongly constrained, based on the data from the nearby universe, $\alpha_T \ne 0$ is still allowed in the past, including at redshifts $\sim 1$ probed by large scale surveys\footnote{There are theoretical arguments disfavouring $c_T > 1$ \cite{Adams:2006sv}, based on the difficulties it creates for embedding the low energy effective dark energy theory into a quantum theory. Due to this, in some of the prior work ({\it e.g} \cite{Perenon:2015sla}), $c_T \le 1$ was enforced as one of the viability conditions. We opt to keep an open mind because gravity is  known to be inconsistent with quantum field theory.}. The agreement with the Big Bang Nucleosynthesis (BBN) requires the Newton's constant $G$ at the time of BBN to be with $10\%$ of the value we measure on Earth, with a similar bound obtained from CMB \cite{Uzan:2010pm}. Aside from these bounds, both $\mu_0$ and $\Sigma_0$ are allowed to vary in the past. However, their values must be consistent with Eqs.~(\ref{mu0_Hornd}) and (\ref{Sigma0_Hornd}) at all times. For example, if one finds that $\mu_0 <1$, which can happen in models of self-accelerating type \cite{Neveu:2013mfa,Barreira:2013xea} because of the increasing $M_*^2(t)$, then one should \emph{not} be observing $\Sigma_0>1$. A situation of this type can only happen if a relatively large positive $\alpha_T$ conspires to change in a very particular way to negate the decrease in $m_0^2/M_*^2$ in $\Sigma_0$, but not in $\mu_0$, which is extremely unlikely.

If the $k$-dependence is detected, so that the $k/a \ll M$ regime can be identified, then \emph{an observation of $\Sigma_0 \ne \mu_0$ or, equivalently, $\gamma_0 \ne 1$, would indicate $c_T \ne 1$}. On the other hand, if one observes  $\Sigma_0=\mu_0$, then one can set $\alpha_T=0$ when examining the bounds in the $k/a \gg M$ limit. 

It is more likely, however, for the transition scale to be outside the observational window. For instance, the $k/a \ll M$ limit would be out of the appropriate range if the mass of the scalar field is set by the horizon scale. If no scale-dependence is detected, then one would have to check the consistency in the small and the large scale limits separately. It may be possible to rule out the large scale regime if one finds that $\Sigma \ne \mu$ at redshifts where $\alpha_T$ is known to be zero, {\it e.g.} from future combined detections of gravity waves and their electromagnetic counterparts \cite{Nishizawa:2014zna}, or CMB B-modes \cite{Amendola:2014wma,Raveri:2014eea}. In such a case, one would abandon the $k/a \ll M$ limit and focus on testing the consistency in the $k/a \gg M$ limit.

\subsection{The small scales limit}

On scales below the Compton wavelength, {\it i.e.} in the limit $k/a \gg M$, the expressions for $\mu$, $\gamma$ and $\Sigma$ become
\ba
\mu_\infty &=&{m_0^2 \over M_*^2}(1+\alpha_T+\beta_\xi^2), \label{mu_inf} \\
\gamma_\infty &=& {1+\beta_B \beta_\xi/2 \over 1+\alpha_T+\beta_\xi^2}, \label{gamma_inf}\\
\Sigma_\infty &=& {m_0^2 \over M_*^2} \left(1+{\alpha_T \over 2} + {\beta_\xi^2 + \beta_B \beta_\xi/2 \over 2} \right). 
\label{Sigma_inf}
\ea
where, following the notation in \cite{Gleyzes:2015rua}\footnote{The definition of $\alpha_B$ in \cite{Gleyzes:2015rua} differs from that in  \cite{Bellini:2014fua} by a factor of $-2$. We use the original definition of \cite{Bellini:2014fua}.}, we defined
\ba
\beta^2_B &=&{2\over c^2_s \alpha} \alpha_B^2 \\
\beta^2_\xi &=& {2 \over c^2_s \alpha} \left[{\alpha_B \over 2}(1+\alpha_T)+\alpha_M-\alpha_T \right]^2 \\
\alpha &=& \alpha_K+{3\over 2}\alpha_B^2 \ ,
\ea
with the expression for the speed of sound of the scalar field perturbations, $c_s^2$, provided in the Appendix.

One can deduce several conclusions from the general forms of $\mu_\infty$ and $\Sigma_\infty$ in (\ref{mu_inf}) and (\ref{Sigma_inf}). Firstly, comparing (\ref{mu_inf}) and (\ref{Sigma_inf}) with (\ref{mu0_Hornd}) and (\ref{Sigma0_Hornd}), we see that one generally should have $\mu_\infty \ge \mu_0$, since $\beta_\xi^2$ is strictly non-negative. This reflects the fact that the scalar fifth force is always attractive and the growth of structure is enhanced on sub-Compton scales as a result. Thus, \emph{if} $k$-dependence is detected, then \emph{finding $\mu_\infty < \mu_0$ would rule out all Horndeski models}. Otherwise, a measurement of $\Sigma_\infty - \Sigma_0$ or $\mu_\infty - \mu_0$ would signal $\beta^2_\xi \ne 0$, which amounts to a detection of a fifth force.

If no scale-dependence is detected, then one must test the small and the large scale regimes one at a time. Evolution of $G_{\rm matter}$ in the $k/a \gg M$ regime has been discussed in great detail in \cite{Perenon:2015sla}. There, it was observed that the time-dependence of $\mu_\infty$ is a combination of the evolution of $\mu_0$ and the enhancement due to the fifth force. As discussed in the previous subsection, it is possible to have $\mu_0 <1$ in models with decreasing $M_*^{-2}$. This can be compensated in $\mu_\infty$ by the fifth force enhancement, potentially giving $\mu_\infty >1$. Indeed, in Covariant Galileon models, $\mu_\infty$  tends to evolve from $<1$ to $>1$ \cite{Neveu:2013mfa,Barreira:2013xea}. The fact that $\Sigma_\infty$ in (\ref{Sigma_inf}) depends on the same parameters as $\mu_\infty$, constrains the differences between the two. For instance, if $\mu_\infty -1$ is of a certain sign at a given epoch, then the sign of $\Sigma_\infty-1$ should be the same, since they share the same $M_*^{-2}$ pre-factor and the corrections they receive from $\alpha_T$ and the fifth force are of the same order.  It is extremely unlikely for $\alpha_T$ and $\beta^2$'s to conspire in just the right way as to negate the effect of $M_*^{-2}$ in $\Sigma_\infty$, but not in $\mu_\infty$. Thus, we conclude that \emph{finding $\mu-1$ and $\Sigma-1$ to be of opposite signs at any redshift or scale would strongly disfavour all Horndeski models.}

Finally, the difference between the values of $\Sigma_\infty$ and $\mu_\infty$ can tell us something about the model. In particular, if $\alpha_T$ is either measured  \cite{Nishizawa:2014zna,Amendola:2014wma,Raveri:2014eea} or assumed to be negligible, then $\Sigma_\infty \ne \mu_\infty$ would amount to a detection of a non-zero $\alpha_M$.

\subsection{The quasi-static approximation}
\label{sec:qsa}

In LCDM, the time derivatives of the metric potentials can be neglected when considering the growth of structure on subhorizon scales. In scalar field models, one can often extend the QSA to also neglect the time derivatives of the scalar field fluctuations on sales below the scalar sound horizon. In models with a canonical form of the scalar field kinetic energy, $c^2_s=1$, but one can have $c^2_s<1$ in more general cases. 

On scales below the sound horizon, due to the pressure support, the scalar field cannot cluster via gravitational instability on its own. Because of this, in minimally coupled models such as quintessence and k-essence, there can be no spatial inhomogeneities in the scalar field on subsonic scales.  In non-minimally coupled models, perturbations in the scalar field still do not cluster on their own on sub-sonic-horizon scales, but, because of the coupling to matter, they are sourced by matter inhomogeneities. Growing matter fluctuations act as a source and the scalar field responds to them. The QSA assumes that the response of the scalar field to matter inhomogeneities is adiabatic, even though there is always dynamics associated with the scalar field response. It can, for example, oscillate about some mean growing inhomogeneity. The key to the validity of QSA is for the dynamical response of the scalar field to not impact the mean adiabatic growth, and also for the oscillations around the mean to be unobservable. 

Above the speed of sound, the scalar field develops its own gravitational instability and no longer traces the matter inhomogeneities. In this case, the QSA is definitely not valid. Thus, restricting the QSA to the sub-sound-hoirzon scales is a necessary condition, which may or may not be sufficient.

In viable non-minimally coupled models with canonical kinetic terms, the QSA tends to hold on sub-horizon scales. Namely, in models with working screening mechanisms of chameleon type \cite{Khoury:2003aq,Hinterbichler:2010es,Damour:1994zq,Brax:2011ja} that have been studied in the literature, the rapid oscillations of the scalar field around the minimum can be ignored \cite{Brax:2012gr,Hojjati:2012rf,Noller:2013wca}. It has also been shown in \cite{Barreira:2012kk} that $c_s^2 \sim 1$ and the QSA holds to a good accuracy in viable Covariant Galileons models \cite{Deffayet:2009wt}. In general, however, the validity of the QSA is model-dependent \cite{Sawicki:2015zya}, and one should try to confirm it before drawing conclusions about specific models based on constraints obtained in terms of the phenomenological functions such as $\mu$ and $\Sigma$.

\section{Case Studies}
\label{sec:examples}

In what follows, we consider two popular sub-classes of Horndeski theories -- generalized models of Brans-Dicke type and Covariant Galileons.

\subsection{Generalized Brans-Dicke models}
\label{sec:gbd}

The Generalized Brans Dicke (GBD) model is described by the action
\be
S = \int d^4x \sqrt{-g} \left[ {\cal L}_{\rm GBD} + {\cal L}_M(g_{\mu \nu},\psi) \right] \ ,
\label{gbd_action}
\ee
with\footnote{One can always absorb $h(\phi)$ into a redefinition of the scalar field. Alternatively, one can set $\Omega=\phi$  and redefine $h$, writing the GBD Lagrangian as $16 \pi G {\cal L}_{\rm GBD} = \phi R - 2 \omega(\phi) \phi^{-1} \partial_\mu \phi \partial^\mu \phi - 2 \Lambda(\phi)$. The original BD model \cite{Brans:1961sx} had constant $\omega$ and $\Lambda$. We opt to keep $h(\phi)$ to make it easier for the reader to convert to different conventions for BD.}
\be
{\cal L}_{\rm GBD} = {\Omega(\phi) \over 16 \pi G} R - {h(\phi)\over 2} \partial_\mu \phi \partial^\mu \phi 
- U(\phi) \ .
\label{gbd_lagrangian}
\ee
It includes the $f(R)$ and chameleon type models, and assumes the weak equivalence principle, i.e. that there exists a Jordan frame metric $g_{\mu \nu}$ to which all matter species (collectively denoted as $\psi$) are minimally coupled. The Einstein's equations in GBD are
\be
\Omega G_{\mu \nu} = 8 \pi G (T^M_{\mu \nu} + T^\phi_{\mu \nu}) + \nabla_\mu \nabla_\nu \Omega-  g_{\mu \nu} \Box \Omega \ ,
\label{gbd_gmunu}
\ee
where $\nabla_\mu$ denotes a covariant derivative, and $T^M_{\mu \nu}$ and $T^\phi_{\mu \nu}$ denote, respectively, the energy-momentum tensor of the matter and the scalar field. One can expand (\ref{gbd_gmunu}) to first order in perturbations and obtain the analogues of the Poisson (\ref{poisson_gr}) and the anisotropy (\ref{shear_gr}) equations in Fourier space. After applying the QSA, they read
\ba
&&\Omega k^2 \Phi = - 4 \pi G a^2 \rho \Delta + {1 \over 2} k^2 \delta \Omega \ ,
\label{poisson_gbd}
\\
&&\Omega k^2 (\Phi - \Psi) = k^2 \delta \Omega \ ,
\label{shear_gbd}
\ea
where we neglected the shear in ordinary matter and radiation. Not that, in GBD, 
\be
T^\phi_{\mu \nu}= h(\phi) \partial_\mu \phi \partial_\nu \phi - g_{\mu \nu} [h(\phi)\partial_\sigma \phi \partial^\sigma \phi/2 + U(\phi)] \ ,
\ee
that has no anisotropic stress ($i \ne j$) component at linear order in perturbations. The effective anisotropic stress appearing on the right hand side of (\ref{shear_gbd}) is due to the conformal factor and is different from the intrinsic scalar field anisotropic stress present only in models with non-trivial $G_4$ or $G_5$ Horndeski terms. It is the intrinsic anisotropic stress that modifies the speed of gravity models, hence $\alpha_T=0$ in GBD.

Combining (\ref{poisson_gbd}) and (\ref{shear_gbd}), we get the analogue of the Poisson equation for the Weyl potential (\ref{weyl_gr}):
\be
k^2(\Phi +\Psi) = -{8\pi G \over \Omega(\phi)} a^2 \rho \Delta \ .
\ee
Comparing it with (\ref{weyl_mg}), we find
\be\label{Sigma_GBD}
\Sigma_{\rm GBD} = \Omega^{-1}\ ,
\ee
{\it i.e.}, \emph{in GBD, $\Sigma$ must be independent of $k$ and is inversely proportional to background value of the conformal factor $\Omega$} which determines the effective background value of the Planck mass.

The other phenomenological functions can be written as \cite{Zhao:2011te,Brax:2012gr}
\ba
\mu &=& \Omega^{-1}({\bar \phi}) [1+\epsilon(k,a)] \\
\label{gamma_GBD}
\gamma &=& {1-\epsilon(k,a) \over 1+ \epsilon(k,a)} \ ,
\ea
where
\be
\epsilon(k,a)={2\beta^2(a) \over 1+M^2(a)a^2/k^2 }
\ee
and $\beta^2$ and $M^2$ denote the coupling and the mass of the scalar field \cite{Brax:2012gr}. Since $\epsilon \ge 0$, one must have $\mu \ge 1$ and $\gamma \le 1$.

Screening mechanisms of chameleon \cite{Khoury:2003aq}, symmetron \cite{Hinterbichler:2010es} and dilaton \cite{Damour:1994zq,Brax:2011ja} type, can suppress the enhancement in $G$ due to the fifth force on scales above the Compton wavelength, but they do not affect the background value of the gravitational coupling. Thus, the value of the Newton's constant that we measure in a screened environment on Earth today must be the same as the current value of the effective Newton's constant $G_{\rm eff} = \mu G$ in the $k\rightarrow 0$ limit. This provides a normalization $\mu(a=1,k=0)$=1, which implies $\Sigma_{\rm GBD} (a=1) = \Omega^{-1}({\bar \phi}_0)=1$, with variations at earlier times constrained by BBN and CMB, as already discussed in the previous Section.

A much stronger bound on changes of $\Omega$ at $z<1$ comes from requiring the screening mechanism to work. As shown in \cite{Wang:2012kj}, this limits variations in the value of the scalar field, implying, in particular, $|\Omega(z=1) - \Omega(z=0)| / \Omega(z=0) \lesssim 10^{-6}$ \cite{Wang:2012kj,Brax:2012gr,Joyce:2014kja}.

The tight restrictions imposed on $\Omega$ in GBD with a scalar field coupling universally to all matter do not apply to models in which the scalar couples with different strengths to baryons and dark matter \cite{Damour:1990tw}. In the case of a non-universal coupling, in addition to the constraints  imposed by measurements of ${\dot G}$, there is a scale-independent bias relating large scale distributions of baryons and dark matter, which, in principle, can be observable \cite{Amendola:2001rc}. Further, there can be significant effects on the mass function of virialized halos \cite{Mainini:2006zj} leading to observable effects on non-linear scales, such as the galaxy satellite abundance, spiral disk formation and apparent baryon shortage, in models that otherwise fit all observations at the level of background and linear perturbations. N-body simulations in non-universally coupled models performed in \cite{Kesden:2006zb,Kesden:2006vz} revealed a relative enhancement of the trailing tidal stream compared to the leading stream in satellite galaxies undergoing tidal disruption. Overall, these effects imply $|\Sigma -1 | \lesssim 0.1$ in non-universally coupled models, but the bounds are expected to become much tighter with future measurements.
 
We conclude that \emph{a measurement of either $\Sigma \ne 1$, $\gamma > 1$ or $\mu < 1$ would rule out all GBD models} with universal coupling to matter, and it would likely also rule out models with non-universal coupling.

\subsection{Covariant Galileons and their generalizations}

Galileons, introduced in \cite{Nicolis:2008in}, are the class of models in which the scalar field Lagrangian is invariant under  the Galilean and the shift symmetry in flat spacetime. In~\cite{Deffayet:2009wt}, this class of models was generalized to a curved spacetime via a covariantization of the Lagrangian. In order to maintain the equations of motion at second order, it was necessary to introduce non-minimal couplings of the scalar field to the curvature, at the cost of loosing the Galilean symmetry. The resulting class of models is a sub-class of Horndeski known as Covariant Galileons. It corresponds to the following choice of the terms in Lagrangian (\ref{S_Horndeski}):
\ba
\label{cov_galileon}
&&K=c_2X,\quad G_3=-\frac{c_3}{M^3}X, \quad G_4= \frac{m_0^2}{2}-\frac{c_4}{M^6}X^2,\nonumber\\
&&G_5=\frac{c_5}{M^9}X^2,
\ea
where $c_2$, $c_3$, $c_4$ and $c_5$ are dimensionless constant, and $M$ is a constant with the dimensions of mass.  

Since there is no cosmological constant, or a quintessence type term, present in (\ref{cov_galileon}), these models must exhibit self-acceleration in order to be viable. This forces the mass of the scalar field to be small, with the corresponding Compton wavelength comparable to the horizon. This implies that all probes of cosmic structure formation that are consistent with the QSA are in the sub-Compton regime. Thus, a detection of any scale-dependence in $\Sigma$ or $\mu$ would rule out all Covariant Galileon models.

Within the Covariant Galileon class, models with $c_2,c_3 \ne 0$ and $c_4=c_5=0$ are referred to as Cubic Galileons (G3), while the cases with $c_4 \ne 0$ and $c_5\ne 0$ are called Quartic (G4) and Quintic (G5) Galileons, respectively. The validity of the QSA in this class of models has been verified on sub-horizon scales in \cite{DeFelice:2010as,Barreira:2014jha}. 

Other than the absence of scale-dependence, the phenomenology of the G4 and G5 models is same as that of general Horndeski in the small scale regime. One can show that all effective functions, $M_*^2$, $\alpha_M$, $\alpha_K$, $\alpha_B$ and $\alpha_T$ can be non-trivial. Thus, in the case of G4 and G5, there is not much to add to the small scale consistency relations discussed in Sec.~\ref{sec:horndeski}.
 
If $G_4=m_0^2/2$ and $G_5=0$, irrespective of the form of $K$ and $G_3$, one has $\alpha_M=\alpha_T=0$ and $M_*^2=m_0^2$. As a consequence, in all such models, which we dub H3, $\beta_\xi=\beta_B/2$. Using this in Eqs.~(\ref{mu0_Hornd})-(\ref{Sigma0_Hornd}) and (\ref{mu_inf})-(\ref{Sigma_inf}), we can immediately find
\be\label{large_scales_cubic}
\mu^{\rm H3}_0=\gamma^{\rm H3}_0=\Sigma^{\rm H3}_0=1
\ee
and
\be\label{small_scales_cubic}
\mu^{\rm H3}_\infty=\Sigma^{\rm H3}_\infty=1+{\beta_B^2 \over 4}, \ \ \gamma^{\rm H3}_\infty=1 \ .
\ee
Thus, an observation of $\mu \ne \Sigma$ or, equivalently, $\gamma \ne 1$, at any scale or any epoch, would rule out all H3 models. This conclusions applies to the G3 models, since they are a sub-class of H3, with the main difference being that H3 can include an explicit Dark Energy driving acceleration, leading to a scale-dependence, while the phenomenology of the G3 models is always in the small-scale regime.

We note that the effective dark energy equation of state in G3 models is, in general, evolving. This makes it challenging to find values of $c_2$ and $c_3$ that simultaneously fit both the background expansion and the growth of cosmological perturbations \cite{Barreira:2012kk,Barreira:2014jha}.

\section{Summary}
\label{sec:summary}

Phenomenological functions $\Sigma$ and $\mu$, parametrizing modified growth of linear perturbations in alternative gravity models, can be well constrained by combining information from the weak lensing and galaxy redshift surveys, together with other cosmological probes. Related tests have already been performed \cite{Simpson:2012ra,Blake:2015vea} using the weak lensing data from CFHTLens \cite{Heymans:2012gg} and RCSLenS \cite{Gilbank:2010zv} combined with growth measurements from WiggleZ \cite{Drinkwater:2009sd} and BOSS \cite{Eisenstein:2011sa}. Measurements of $\Sigma$ and $\mu$ will become significantly more accurate  \cite{Hojjati:2011xd,Asaba:2013mxj} with future surveys, such as Euclid and LSST. Given these prospects, we asked in this paper if there are trends and consistency relations that must be respected by $\Sigma$ and $\mu$ within the general paradigm of viable scalar-tensor theories of gravity of Horndeski type. We identified several such conditions which could help to rule out large sub-classes of Horndeski, irrespective of the particular parametric forms of $\Sigma$ and $\mu$. They are presented as a flow chart diagram in Fig.~\ref{fig:chart}.

For example, an observation of $\mu \ne \Sigma$ at any scale or any epoch, would rule out Cubic Galieons and all Horndeski models of H3 type, {\it i.e.} those with $G_4=m_0^2/2$, $G_5=0$, and arbitrary $K$ and $G_3$. An observation of scale-dependence in any of the phenomenological functions would rule out Covariant Galileons and other models exhibiting self-acceleration. An observation of $\Sigma \ne 1$ would rule out all models with canonical kinetic terms.

If a scale-dependence is detected in either $\Sigma$ or $\mu$, then the difference between their large scale and small scale limiting values amounts to a detection of a fifth force. Also, one generally should have $\Sigma_\infty > \Sigma_0$ and $\mu_\infty > \mu_0$, since the force mediated by the scalar is attractive. If no scale-dependence is detected, it is possible to figure out if you are probing the large or the small scale regime, since the difference between $\Sigma_0$ and $\mu_0$ can only be due to $\alpha_T \ne 0$. The bound on the speed of gravity waves at cosmological redshifts may eventually become available, {\it e.g.} from electromagnetic counterparts of binary mergers \cite{Nishizawa:2014zna}, or CMB B-modes \cite{Amendola:2014wma,Raveri:2014eea}.

In all cases, the difference between the values of $\Sigma$ and $\mu$ can tell us something about the model. For example, if $\alpha_T$ is either measured or assumed to be negligible, an observation of $\Sigma \ne \mu$ would rule out the large-scale regime, and would amount to a detection of a non-zero $\alpha_M$, while a measurement of $\mu-1$ and $\Sigma-1$ to be of opposite signs on any scale would strongly disfavour all Horndeski models. It is interesting to note that the best fit values of the phenomenological functions derived by Planck Collaboration in Ref.~\cite{Ade:2015rim} indicate $\mu<1$ and $\Sigma>1$. The statistical significance of the departure from LCDM is too low to be a cause for concern. However, it serves as an illustration that, if such values were to hold up, they would effectively rule out all Horndeski models.

Our results demonstrate the utility of the ($\Sigma,\mu$) approach to testing gravity on cosmological scales, and how far reaching conclusions about alternative gravity theories can be derived independently from details of their parametrizations.

\acknowledgments

We thank Tessa Baker, David Langlois, Michele Mancarella, Louis Perenon, Federico Piazza, Marco Raveri, Dani Steer, Filippo Vernizzi and Alex Zucca for useful discussions. LP is supported by NSERC and is grateful to APC at the University of Paris 7 for hospitality. AS acknowledges support from The Netherlands Organization for Scientific Research (NWO/OCW), and from the D-ITP consortium, a program of the Netherlands Organisation for Scientific Research (NWO) that is funded by the Dutch Ministry of Education, Culture and Science (OCW).  AS
also thanks the COST Action (CANTATA/CA15117), supported by COST (European Cooperation in Science and Technology).
\appendix

\section{}
\label{sec:eom-eft}

The equations of motion for perturbations in a general scalar-tensor theory in the quasi-static limit can be written as \cite{Bloomfield:2012ff}
\ba
A_1 {k^2 \over a^2} \Phi + A_2 {k^2 \over a^2} \pi + A_3 {k^2 \over a^2} \Psi = -\rho \Delta ,
\label{eom_poisson}
\\
B_1 \Psi + B_2 \Phi + B_3 \pi = 0,
\label{eom_anisotropy}
\\
C_1 {k^2 \over a^2} \Phi + C_2 {k^2 \over a^2} \Psi + \left(C_3 {k^2 \over a^2} + C_\pi  \right) \pi = 0,
\label{eom_pi}
\ea
where we stick to the notation of \cite{Bloomfield:2012ff}. Expressed in terms of the functions in the EFT action (\ref{EFT_action}), with the spatial curvature set to zero, the coefficients are
\ba
A_1 &=& 2m_0^2\Omega+4{\hat M}^2 \nonumber \\
A_2 &=& -m_0^2 \dot{\Omega}-\bar{M}^3_1+2H\bar{M}^2_3+4H\hat{M}^2 \nonumber \\
A_3 &=& -8m_2^2 \nonumber \\
B_1 &=& -1-{2\hat{M}^2 \over m_0^2\Omega} \nonumber \\
B_2 &=& 1 \nonumber \\
B_3 &=& -{\dot{\Omega} \over \Omega} + {\bar{M}^2_3 \over m_0^2 \Omega} \left( H+ {2 \dot{\bar M}_3 \over \bar{M}_3} \right) \nonumber \\
C_1 &=& m_0^2 \dot{\Omega} +2H\hat{M}^2+4\hat{M}\dot{\hat M} \nonumber \\
C_2 &=& -{m_0^2\over 2} \dot{\Omega} - {1\over 2} \bar{M}_1^3 - {3\over 2} H \bar{M}_2^2 - {1\over 2} H \bar{M}_3^2 + 2H\hat{M}^2 \nonumber \\
C_3 &=& c-{1\over 2} (H+\partial_t)\bar{M}_1^3 + \left({k^2\over 2a^2}-3\dot{H}\right) \bar{M}_2^2 \nonumber \\
&+& \left( {k^2\over 2a^2} -\dot{H}\right) \bar{M}_3^2 + 2(H^2+\dot{H}+H\partial_t) \hat{M}^2 \nonumber \\
C_\pi &=& {m_0^2\over 4}\dot{\Omega}\dot{R}^{(0)}-3c\dot{H} + {3\over 2} ( 3H\dot{H} +\dot{H} \partial_t+\ddot{H}) \bar{M}_1^3 
\nonumber \\
&+& {9 \over 2} \dot{H}^2 \bar{M}_2^2 +  {3 \over 2} \dot{H}^2 \bar{M}_3^2 
\label{M2_Hornd}
\ea
Under the QSA, the phenomenological functions $\mu$, $\gamma$ and $\Sigma$ can be written as
\ba
4\pi G \mu = {\mu \over 2m_0^2}&=&{f_1+f_2 \ a^2/k^2 \over f_3+f_4 \ a^2/k^2},
\label{eq:mu-eft} \\
\gamma &=& {f_5+f_6 \ a^2/k^2 \over f_1+f_2 \ a^2/k^2},
\label{eq:gamma-eft} \\
8\pi G \Sigma={\Sigma \over m_0^2} &=& {f_1+f_5+(f_2+f_6) a^2/k^2 \over f_3+f_4 \ a^2/k^2},
\label{eq:Sigma-eft}
\ea
where
\ba
f_1&=&B_2C_3-C_1B_3 \nonumber \\
f_2&=&B_2C_\pi \nonumber \\
f_3&=&A_1(B_3C_2-B_1C_3)+A_2(B_1C_1-B_2C_2) \nonumber \\
&+&A_3(B_2C_3-B_3C_1) \nonumber \\
f_4&=& (A_3B_2-A_1B_1)C_\pi \nonumber \\
f_5&=&  B_3C_2-B_1C_3 \nonumber \\
f_6&=&  -B_1 C_\pi
\label{eq:f_i}
\ea
In the case of Horndeski theories, with $m_2^2=0$ and $2{\hat M}^2 = {\bar M}^2_2 =-{\bar M}^2_3$, the $k$-dependence in $C_3$ disappears, and $f_1$,...,$f_5$ are functions of time only \cite{Silvestri:2013ne}.

The functions appearing in the effective action given by Eq.~(\ref{alpha_action}) are related to solutions of Horndeski theories via \cite{Bellini:2014fua}
\ba
&&M_*^2 = 2[G_4-2XG_{4X}+XG_{5\phi}-{\dot \phi}HXG_{5X}] \\
&&HM_*^2 \alpha_M = {d M_*^2 \over dt} \\
&&M_*^2 \alpha_T = 2X[2G_{4X}-2G_{5\phi}-(\ddot{\phi} - H\dot{\phi}) G_{5X}] \\ \nonumber
&&HM_*^2 \alpha_B = 2\dot{\phi} [XG_{3X}-G_{4\phi}-2XG_{4\phi X}] \\ \nonumber
&& \ \ \ +8XH(G_{4X}+2XG_{4XX}-G_{5\phi}-XG_{5\phi X}) \\
&& \ \ \ + 2\dot{\phi}XH^2[3G_{5X}+2XG_{5XX}] \\ \nonumber
&&HM_*^2 \alpha_K = 2X[K_X+2XK_{XX}-2G_{3\phi}-2XG_{3\phi X}] \\
&& \ \ \ + 12\dot{\phi}XH[G_{3X}+XG_{3XX}-3G_{4\phi X}-2XG_{4\phi XX}] \nonumber \\
&& \ \ \ +12XH^2[G_{4X}+8XG_{4XX} + 4X^2 G_{4XXX}] \nonumber \\
&& \ \ \ -12XH^2[G_{5\phi} +5XG_{5\phi X}+2X^2G_{5\phi XX}] \nonumber \\
&& \ \ \ +4\dot{\phi}XH^3[3G_{5X}+7XG_{5XX}+2X^2G_{5XXX}]
\ea
They are related to the functions appearing in the EFT action (\ref{EFT_action}) via \cite{Bellini:2014fua}
\ba
M_*^2 &=& m_0^2\Omega + \bar{M}^2_2 
\label{eq:M*} \\ 
HM_*^2 \alpha_M &=& m_0^2\dot{\Omega} + \dot{\bar{M}}^2_2 \\
M_*^2 \alpha_T &=&  -\bar{M}^2_2 
\label{eq:alphaT} \\
HM_*^2 \alpha_B &=& -m_0^2\dot{\Omega} - \bar{M}^3_1 \\
HM_*^2 \alpha_K &=& 2c+4M_2^4 \ .
\ea
The speed of sound of the scalar field perturbations is given by
\ba
\nonumber
c_s^2&=&{2\over \alpha} \Big[ \left(1-{\alpha_B \over 2} \right) \Big( \alpha_M-\alpha_T+ {\alpha_B \over 2 }(1+\alpha_T) -{\dot{H} \over H^2} \Big) \\
&+&{\dot{\alpha}_B \over 2H} - {3 \over 2} \Omega_M  \Big] \ .
\ea
\bibliography{sigmapaper}

\end{document}